\documentclass[twocolumn,reprint,amsmath,amssymb,aps]{revtex4-2}
\usepackage{graphicx}
\usepackage{dcolumn}
\usepackage{bm}
\usepackage{float}
\usepackage{xcolor}

\begin{document}

\preprint{APS/123-QED}

\title{Wigner-negative states in the steady-state emission of a two-level system driven by squeezed light}

\author{Miriam.~J. Leonhardt}
\author{Scott Parkins}
\affiliation{Dodd-Walls Centre for Photonic and Quantum Technologies, New Zealand}
\affiliation{Department of Physics, University of Auckland, Auckland 1010, New Zealand}

\date{\today}

\begin{abstract}
Propagating modes of light with negative-valued Wigner distributions are of fundamental interest in quantum optics and represent a key resource in the pursuit of optics-based quantum information technologies. Most schemes proposed or implemented for the generation of such modes are probabilistic in nature and rely on heralding by detection of a photon or on conditional methods where photons are separated from the original field mode by a beam splitter. We demonstrate theoretically, using a cascaded-quantum-systems model, the possibility of deterministic generation of Wigner-negativity in temporal modes of the steady-state emission of a two-level system driven by finite-bandwidth quadrature-squeezed light. Optimal negativity is obtained for a squeezing bandwidth similar to the linewidth of the transition of the two-level system. While the Wigner distribution associated with the incident squeezed light is Gaussian and everywhere positive, the Wigner functions of the outgoing temporal modes show distinct similarities and overlap with a superposition of displaced squeezed states.
\end{abstract}

\maketitle

\section{Introduction} 
In recent years there has been significant interest in techniques for the production of propagating modes of light with negativity in their associated Wigner quasi-probability distributions \cite{kitten1,Chiral&Interferometry,GKP_states,BS_kitten,SS-feeback,1_atom_2018,1_atom_2019,Wig_Neg_Expt,Wig_Neg_KPO,2photons,Review_QSL,QSL_Furusawa}. 
Negativity in the Wigner distribution is an unequivocal indicator of non-classicality \cite{HUDSON,WN_volume} and such modes represent a key resource in quantum computation with continuous variables \cite{CV_quantum_computation,WN_CV_computation}. 
Currently, many schemes for the generation of non-classical states of light, in particular Wigner-negative states, require heralding \cite{QSL_Furusawa,2photons} or conditional measurements \cite{GKP_states,BS_kitten,Review_QSL}, both of which are inherently probabilistic, or they may depend on transient dynamics of a system \cite{Chiral&Interferometry}, resulting in low generation rates of the desired states. Alternatively, one may consider generating steady-state Wigner-negative light by using feedback for stabilisation \cite{SS-feeback}. In this vein, however, an arguably much simpler method was recently suggested by Strandberg {\em et al}. \cite{1_atom_2018,1_atom_2019} involving a coherently-driven, two-level system. Their scheme extracts Wigner-negative temporal modes from the steady-state output field of the coherently-driven, two-level atom, and the theoretical predictions have indeed been experimentally verified using a circuit QED set-up \cite{Wig_Neg_Expt}.

In this work, we extend the work of \cite{1_atom_2018,1_atom_2019} to a two-level system driven continuously with {\em squeezed light} instead of coherent light. We demonstrate numerically the unconditional generation of Wigner function negativity in appropriately defined temporal modes of the backwards (or reflected) emission of a two-level system driven by finite-bandwidth quadrature-squeezed light produced by a degenerate parametric amplifier. 
Furthermore, the Wigner function displays a richer structure of negativity than the coherent-driving case and, in fact, the state generated shows intriguing and quantifiable similarities with a squeezed Schr\"odinger cat state \cite{2photons}.

We use a cascaded systems model \cite{Gardiner1993,CascadeHoward} together with the input-output theory for quantum pulses introduced by Kiilerich and Mølmer \cite{Capture_Cavity1,cc_theory} to investigate temporal modes of the propagating output field. 

The paper is organised as follows. In Sec.~\ref{sec:model}, we present the model used to describe this cascaded quantum system, and the method in which we calculate temporal modes of the output field of the system. In Sec.~\ref{sec:results}, we show that Wigner negativity exists in the temporal mode states of the steady-state output of this system, and we investigate how the negativity content of the Wigner distribution of these states depends on key system parameters, as well as on the elements of the density matrices describing the states of the modes. We show, for example, that maximum negativity is obtained for a squeezing bandwidth similar to the linewidth of the two-level transition, making this an intrinsically non-Markovian problem with respect to the behavior of the emitter \cite{Ritsch1988,Parkins1988}. We also explore how the purity of the temporal mode state and the degree of excitation of the two-level system change with system parameters, and demonstrate an anticorrelation between these two quantities. In Sec.~\ref{sec:sq_kittens} we investigate intriguing similarities between the resulting Wigner-negative temporal mode states and squeezed Schr\"odinger cat states. Finally, in Sec. \ref{sec:conclusion}, we summarise our findings.
\section{Model}\label{sec:model} 
\subsection{Cascaded Quantum Systems}
\begin{figure}[t]
    \centering    \includegraphics[width=0.4\textwidth]{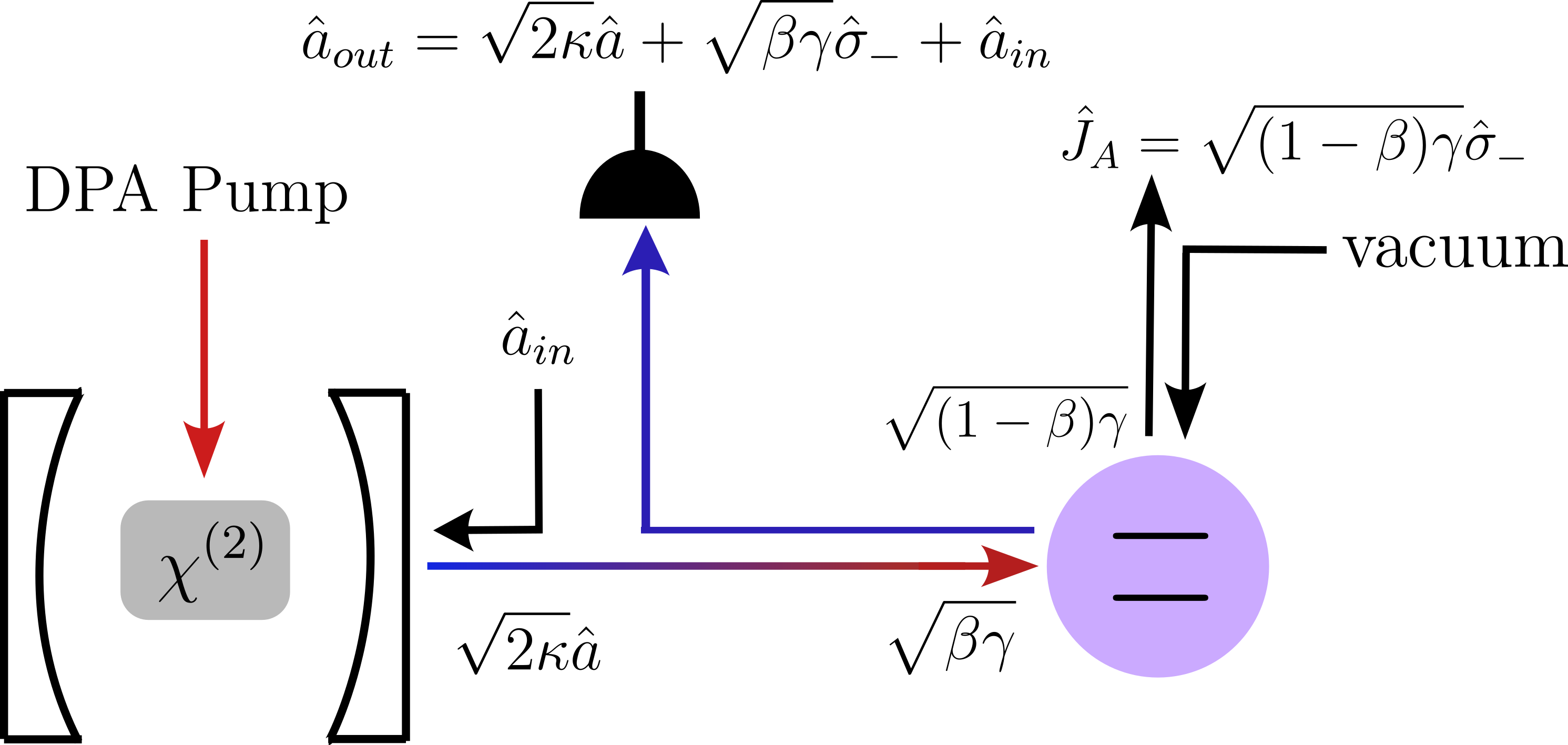}
    \caption{Schematic of the physical system under consideration: the steady-state output of a degenerate parametric amplifier (DPA) is used to drive a two-level system (TLS). We model this system as a cascaded system, where the coupling between the DPA and TLS is unidirectional. The DPA is modeled as a one-sided optical cavity containing a medium with a second order nonlinear susceptibility. The TLS has two decay channels, one coupled to the reservoir that connects the DPA with the TLS, and one that accounts for all other free-space emission. We observe temporal modes of the combined steady-state output of the DPA and the TLS.}
    \label{fig:schematic}
\end{figure}

In our setup the two-level system (TLS) is driven by squeezed light produced in the steady-state output field of a degenerate parametric amplifier (DPA) \cite{InputOutput_DPA}, as depicted in Fig.~\ref{fig:schematic}. 
Using the cascaded-systems formalism of \cite{Gardiner1993,CascadeHoward}, this setup is described by the Hamiltonian (in a frame rotating at the DPA carrier frequency)
\begin{align}\label{eq:sys_ham}
    \hat{H}_{S}&=i\hbar\frac{\lambda}{2}(\hat{a}^{\dagger^2}-\hat{a}^2)+\hbar\Delta_A\hat{\sigma}_+\hat{\sigma}_-
    \nonumber
    \\
    & ~~~+\frac{i\hbar}{2}\sqrt{2\kappa\beta\gamma}(\hat{a}^\dagger\hat{\sigma}_--\hat{a}\hat{\sigma}_+),
\end{align}
where $\hat{a}$ ($\hat{a}^\dagger$) is the annihilation (creation) operator for the cavity mode of the DPA, $\kappa$ is the linewidth (HWHM) of the DPA cavity mode, and $\lambda$ is proportional to the second-order nonlinear susceptibility, $\chi^{(2)}$, of the parametric medium and to the strength of the pump field driving it. 
The DPA is operated (resonantly) below threshold and the maximum degree of squeezing in its output field is determined by the ratio $\lambda/\kappa~(<1)$.
The lowering (raising) operator of the TLS is $\hat{\sigma}_-$ ($\hat{\sigma}_+$), and $\Delta_A$ is the detuning of the transition frequency of the TLS from the DPA carrier frequency  (we generally assume $\Delta_A=0$). 
The total emission rate of the TLS is $\gamma$, and the beta-factor, $\beta$ ($0\leq\beta\leq 1$), determines the fraction of this emission coupled into the channel through which the TLS is driven. The remaining portion of the emission can be used to model imperfect coupling through decay into free-space modes. 

Using the input-output formalism \cite{InputOutput_DPA,Input_output_SDE}, the output field we focus on is given by
\begin{equation}\label{eq:J_out_DPA}  
\hat a_{out}(t) = \hat{J}_{out}(t) + \hat a_{in}(t),
\end{equation}
where $\hat{J}_{out}=\sqrt{2\kappa}\hat{a}+\sqrt{\beta\gamma}\hat{\sigma}_-$ and $\hat a_{in}(t)$ is the (vacuum) input field to the DPA. That is, we consider the backwards (or reflected) emission,
which contains contributions from both the DPA and the TLS.

Considering the environment as a zero-temperature reservoir, this open cascaded quantum system is described by the master equation
\begin{align}
    \frac{d}{dt}\hat{\rho}&=\frac{1}{i\hbar}[\hat{H}_S,\hat{\rho}]+\frac{1}{2}\mathcal{D}[\hat{J}_{out}]\hat{\rho}+\frac{1}{2}\mathcal{D}[\hat{J}_A]\hat{\rho},\label{eq:ME_lindblad}
\end{align}
where $\hat\rho$ is the density operator of the whole system, $\mathcal{D}[\hat{\cal O}]\hat{\rho}\equiv 2\hat{\cal O}\hat\rho\hat{\cal O}^\dagger-\hat{\cal O}^\dagger\hat{\cal O}\hat\rho-\hat\rho\hat{\cal O}^\dagger\hat{\cal O}$, and $\hat{J}_A=\sqrt{(1-\beta)\gamma}\hat{\sigma}_-$.
We use (\ref{eq:ME_lindblad}) to determine the steady state of the system and its output. 
\subsection{Temporal Modes of the Output Field}
The output field defined by Eq.~\eqref{eq:J_out_DPA} is a propagating field and thus corresponds to a continuum of modes. We extract a single temporal mode from the output field, and then determine the Wigner distribution for this temporal mode. The temporal mode is defined through the filtered output field operator,
\begin{equation}\label{eq:TM_def}
    \hat{A}_{out,v}=\int_0^\infty v(t)\hat{a}_{out}(t)dt,
\end{equation}
where $v(t)$ is the filter function determining the temporal shape of the wavepacket \cite{TM}. The filter function obeys the normalisation condition $\int_0^\infty|v(t)|^2dt=1$, so that  $[\hat{A}_{out,v},\hat{A}_{out,v}^\dagger]=1$. In this work we choose to apply Gaussian filter functions to the output field,
\begin{align}\label{eq:Gaussian}
    v(t)=\left(\frac{8}{\pi\tau^2}\right)^{\frac{1}{4}}
    \exp \left[ - \left(\frac{t-t_0}{\tau /2}\right)^2\right] ,
\end{align}
where $T_v=\tau\sqrt{\ln (2)}$ is the temporal full width at half maximum (FWHM) of the filter function, and $t_0$ is chosen so that it shifts the centre of $v(t)$ to the middle of the filtering interval, ensuring the normalisation over $[0,\infty)$ is correct to sufficient precision. The choice of a Gaussian filter is based on the desire for a simple filter function that is smooth and also maximises (at least approximately) the negativity content of the Wigner distributions of the temporal mode. 
Consistent with 
Refs.~\cite{Wig_Neg_Expt, Wig_Neg_KPO}, we find that box-car filters with comparable widths yield similar results, but Gaussian filters consistently produce temporal modes with noticeably larger negativity content in their Wigner distributions.

To extract these temporal modes in the numerical simulations using QuTiP \cite{qutip2}, we make use of the input-output formalism for quantum pulses \cite{Capture_Cavity1,cc_theory}. This formalism shows that the quantum state contained in a temporal mode of pulse shape $v(t)$ can be ``captured'' in the quantized mode of a virtual cavity that has a time-dependent coupling to the incident field of the form
\begin{equation}\label{eq:cc_coupling}
    g_v(t)=\frac{-v^*(t)}{\sqrt{\int_0^t|v(t')|^2dt'}}.
\end{equation}
This also works for a driven quantum system such as ours, which produces a continuous output. 
The virtual cavity is included in the model by again using the cascaded systems formalism to describe the 
unidirectional coupling of the output field $\hat a_{out}(t)$ to the virtual cavity.

In practice, homodyne measurements combined with maximum likelihood estimation \cite{Iterative_method,MLE_Hradil,EMU} would be used 
\onecolumngrid
\begin{center}
\begin{figure}[t!]
        \centering
        \includegraphics[width=0.97\textwidth]{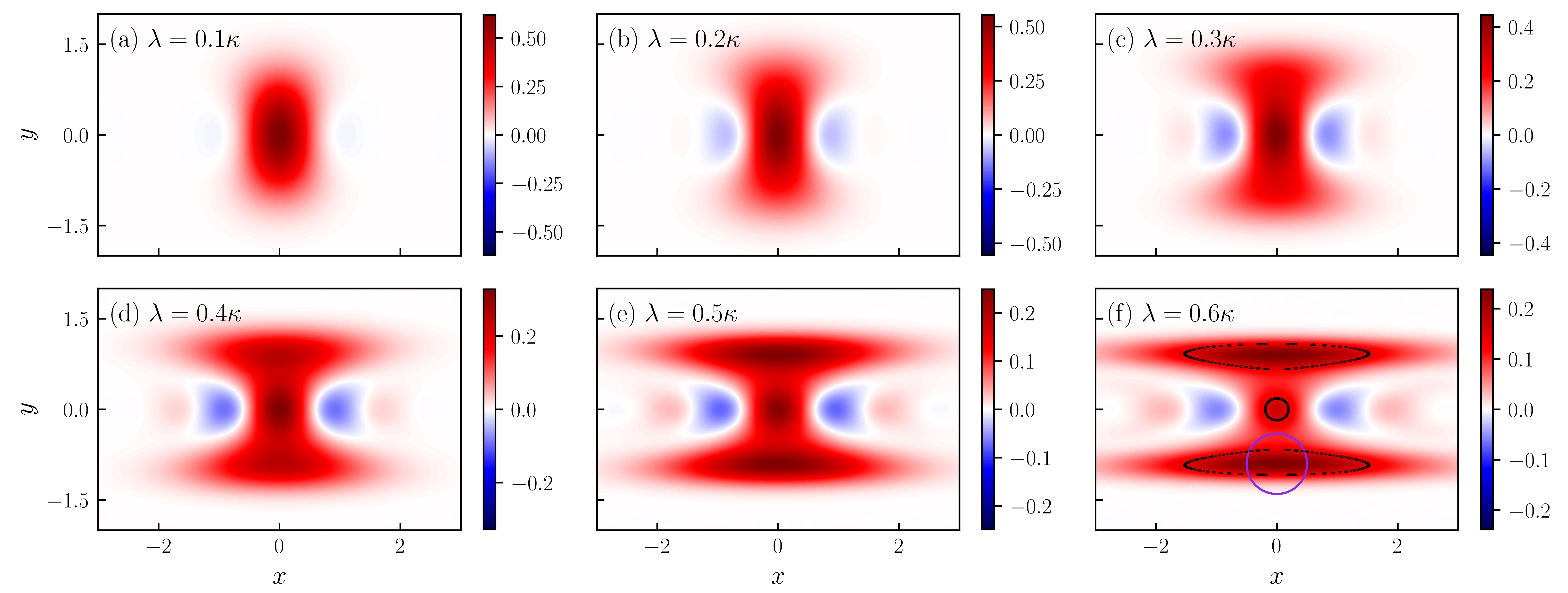}
        \caption{Wigner distributions for Gaussian temporal modes of the output field $\hat a_{out}(t)$ with varying degrees of squeezing $(\lambda)$. The FWHM's of the filter function are $T_v = \{12.0, 10.4, 8.8, 6.4, 5.6, 5.0\}\kappa^{-1}$, respectively, while $\gamma = 0.5\kappa$ and $\beta =1$. The minimum values of the Wigner distributions are $\{-0.014,-0.069,-0.097,-0.092,-0.076,,-0.056\}$, respectively. Note the different scales of the colour bars. In (f), the solid purple circle represents the error circle ($e^{-1/2}$ contour) of a coherent state, centred at the lower maximum of the temporal mode Wigner distribution. The equivalent contour for the temporal mode is shown by the dashed black curves. Note that all Wigner distributions have been calculated with the quadrature convention $\hat{x}=1/2(\hat{a}+\hat{a}^\dagger)$.}
        \label{fig:Wigner_functions}
\end{figure}
\end{center}
\twocolumngrid
\noindent in an experiment to determine the state of the temporal mode \cite{Wig_Neg_Expt}. Such measurements can be simulated using quantum trajectory theory \cite{Howard2,1_atom_2019} and we have also used this method to confirm the results presented here.

\section{Results}\label{sec:results}
\subsection{Wigner Function Negativity}
In Fig.~\ref{fig:Wigner_functions} we show the Wigner distributions of Gaussian temporal modes of the output field $\hat a_{out}(t)$ for varying values of the effective DPA pump strength $\lambda$. 
Here, $\lambda$ takes on values between $0.1\kappa$ and $0.6\kappa$. The maximum degree of squeezing in the output field produced by the DPA increases with $\lambda$ and occurs at the DPA carrier frequency. Using the input-output formalism ~\cite{InputOutput_DPA,Input_output_SDE}, this maximum squeezing ranges from 1.74~dB to 12.0~dB of noise reduction below the vacuum level for these choices of $\lambda$ \cite{dBformula}. In all instances the states are clearly non-classical, as evidenced by the negativity in the Wigner distributions. The Wigner distributions of these states are distinct from those seen with coherent driving in Refs.~\cite{1_atom_2018,1_atom_2019}; one now observes {\em two} pronounced regions of negativity. 
\begin{figure}[t!]
    \centering    \includegraphics[width=0.45\textwidth]{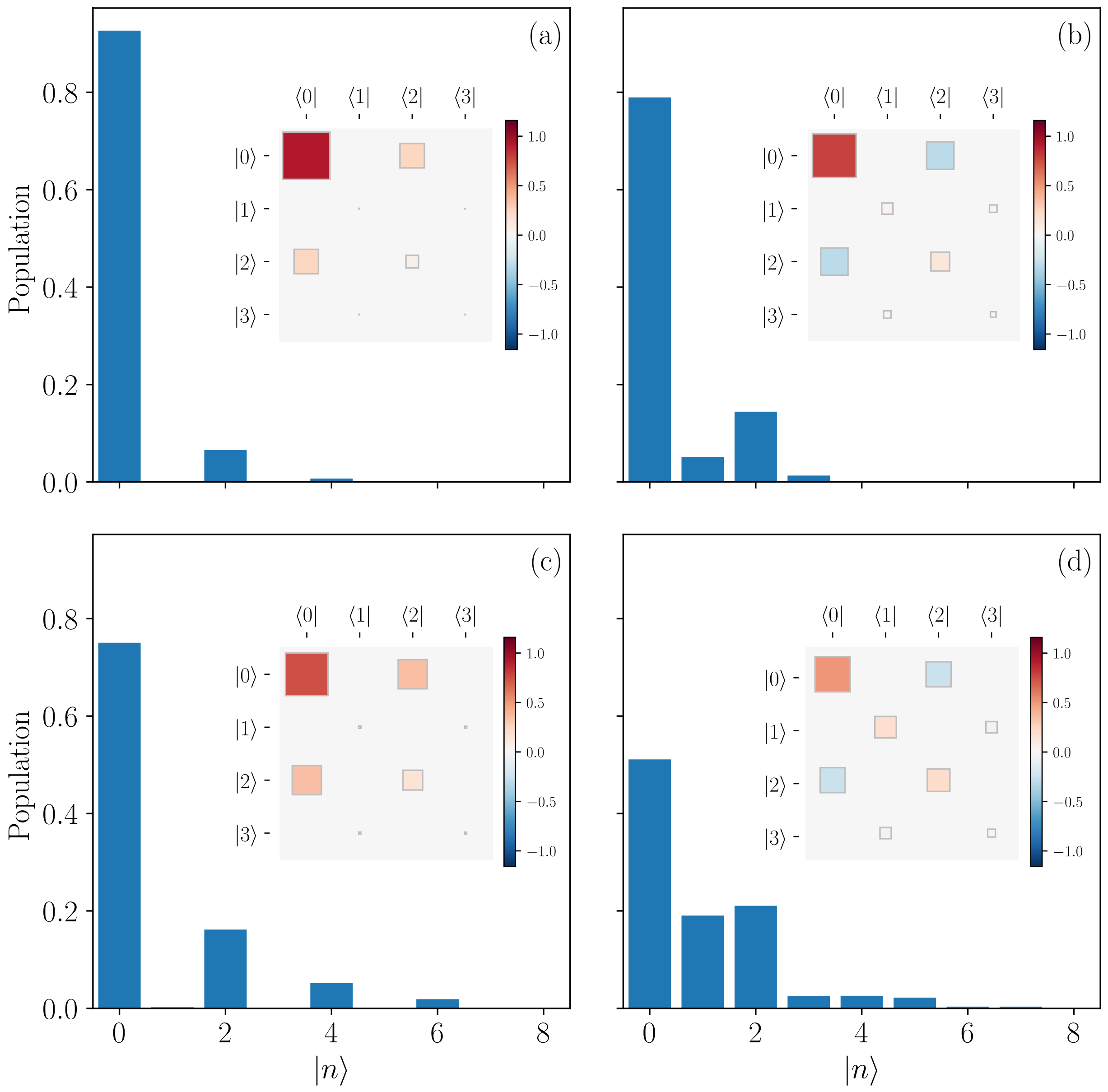}
    \caption{Fock state populations of the Gaussian temporal modes of the output field for different squeezing strengths and different couplings ($\beta$), with fixed $\gamma=0.5\kappa$. The top row (a and b) corresponds to a squeezing strength of $\lambda=0.2\kappa$ and a temporal mode with $T_v=10.4\kappa^{-1}$, while the bottom row (c and d) corresponds to $\lambda=0.4\kappa$ with $T_v=6.4\kappa^{-1}$. The left column (a and c) corresponds to $\beta=0$ when only light from the DPA contributes to the temporal mode, and the right column (b and d) corresponds to $\beta=1$. The insets show the Hinton plot of the corresponding density matrices; the size of the squares, as well as the darkness of the color indicates the magnitude of the element, and the color of the square indicates the sign of the
corresponding element.}
    \label{fig:hinton}
\end{figure}
This is related to the fact that the temporal mode states of Fig.~\ref{fig:Wigner_functions} are absent of coherence between Fock states differing by an odd number of photons (e.g., $\langle0|\hat{\rho}_v|1\rangle$, where $\hat\rho_v$ denotes the density matrix of the temporal mode), reflecting the similar absence of such coherence in the squeezed light driving the TLS. 
Instead, the most significant off-diagonal terms in the density matrix are $\langle0|\hat{\rho}_v|2\rangle$ and $\langle1|\hat{\rho}_v|3\rangle$, as illustrated in Fig.~\ref{fig:hinton}, where Fock state populations and Hinton plots of the temporal mode density matrices are shown for two different squeezing strengths (corresponding to Figs.~\ref{fig:Wigner_functions}(b) and (d), respectively), both with and without coupling of the TLS to the incident squeezed light ($\beta=1$ and $\beta=0$, respectively). 
In particular, while the states of temporal modes extracted directly from the DPA output field ($\beta=0$, Figs.~\ref{fig:hinton}(a) and (c)) are dominated by even-photon-number states and their associated coherences, interaction with the TLS leads to significant contributions from odd-photon-number states and coherence dominated by $\langle0|\hat{\rho}_v|2\rangle$ {\it and} $\langle1|\hat{\rho}_v|3\rangle$ ($\beta=1$, Figs.~\ref{fig:hinton}(b) and (d)). The density matrix elements $\langle0|\hat{\rho}_v|2\rangle$ and $\langle1|\hat{\rho}_v|3\rangle$
each contribute, for a given radius in quadrature phase space, two regions of negativity to the Wigner distribution, in contrast to the single region associated with a finite value of the coherence $\langle0|\hat{\rho}_v|1\rangle$. 

The signs of $\langle0|\hat{\rho}_v|2\rangle$ and $\langle1|\hat{\rho}_v|3\rangle$ are also important in determining the Wigner distribution. Importantly, after interaction of the squeezed light with the TLS, the sign of the coherence $\langle0|\hat{\rho}_v|2\rangle$ in the temporal mode state has switched compared to the case where there was no interaction ($\beta=0$). This means that the two regions of negativity associated with this coherence are rotated around the origin of phase space by $\pi/2$ and they now contribute to the Wigner-negative regions on the left- and right-hand side of the central maximum, rather than to the squeezing observed in the $y$-quadrature of the temporal mode extracted solely from the DPA output field. 
We emphasize that this sign change, and the contributions from the single-photon Fock state and its associated coherences, all resulting from the interaction of the DPA output field with the TLS, are  central to producing the Wigner negativity seen in Fig.~\ref{fig:Wigner_functions} \cite{MSc}. 

Fig.~\ref{fig:hinton} also highlights that, for temporal modes extracted solely from the DPA output field, odd-photon-number contributions arising from the possibility of ``missing'' one photon of a pair in the finite duration of the temporal mode (as in \cite{Wig_Neg_KPO}) are essentially negligible in comparison with those resulting from the interaction of the squeezed light with the TLS. 
That is, the TLS clearly enhances the temporal ``break-up'' of photons within a pair, which can be attributed quite naturally to the fact that the TLS can only absorb and subsequently emit one photon at a time. Over the finite duration of the temporal mode, this obviously enhances the probability of contributions from odd-photon-number states to the state of the mode. 

\begin{figure}[t!]
    \centering    \includegraphics[width=0.45\textwidth]{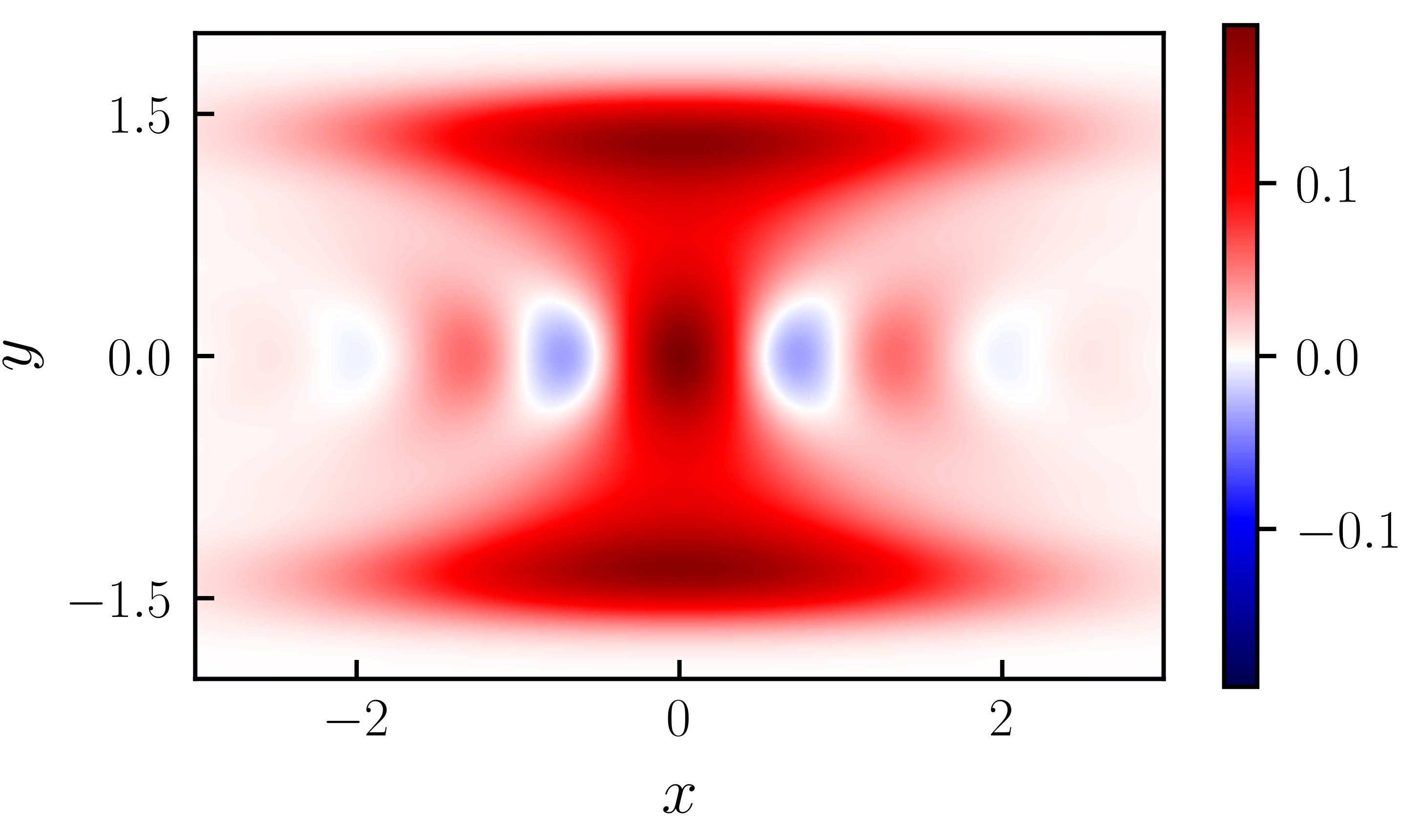}
    \caption{Wigner distribution for a Gaussian temporal mode with $T_v=12.0\kappa^{-1}$, with $\lambda=0.5\kappa$, $\gamma=0.5\kappa$ and $\beta=1$. Four distinct regions of Wigner-negativity are visible.}
    \label{fig:4_lobes}
\end{figure}

The two distinct regions of negativity seen in Fig. \ref{fig:Wigner_functions} are typical of temporal modes extracted from the steady-state output of this system. However, for certain parameter choices it is possible to observe additional regions of Wigner-negativity, as illustrated in Fig.~\ref{fig:4_lobes}. 
This requires temporal modes with larger $T_v$ and sufficiently strong squeezing ($\lambda\geq0.4\kappa$). Note that in our simulations the squeezing parameter has been limited to $\lambda\lesssim 0.6\kappa$, as the numerical resources (basis sizes) required to perform accurate calculations for very strong squeezing become extremely challenging. 

In the presence of an additional atomic decay channel in the model, corresponding to the decay operator $\hat{J}_A$ (see Fig. \ref{fig:schematic}), it is also possible to consider temporal modes in the steady-state output field of the TLS, where there are no direct contributions from the output field of the DPA. However, in the parameter regimes explored, temporal modes in this free-space output do not exhibit Wigner-negativity. This is related to the fact that, while a small value of $\beta$ is desirable to allow substantial decay into this channel, a value of $\beta$ close to 1 is needed for the TLS to be driven sufficiently strongly by the output of the DPA to produce a significant radiated field.

\subsection{Negative Volume}
\begin{figure}[t!]
  \includegraphics[width=0.23\textwidth]{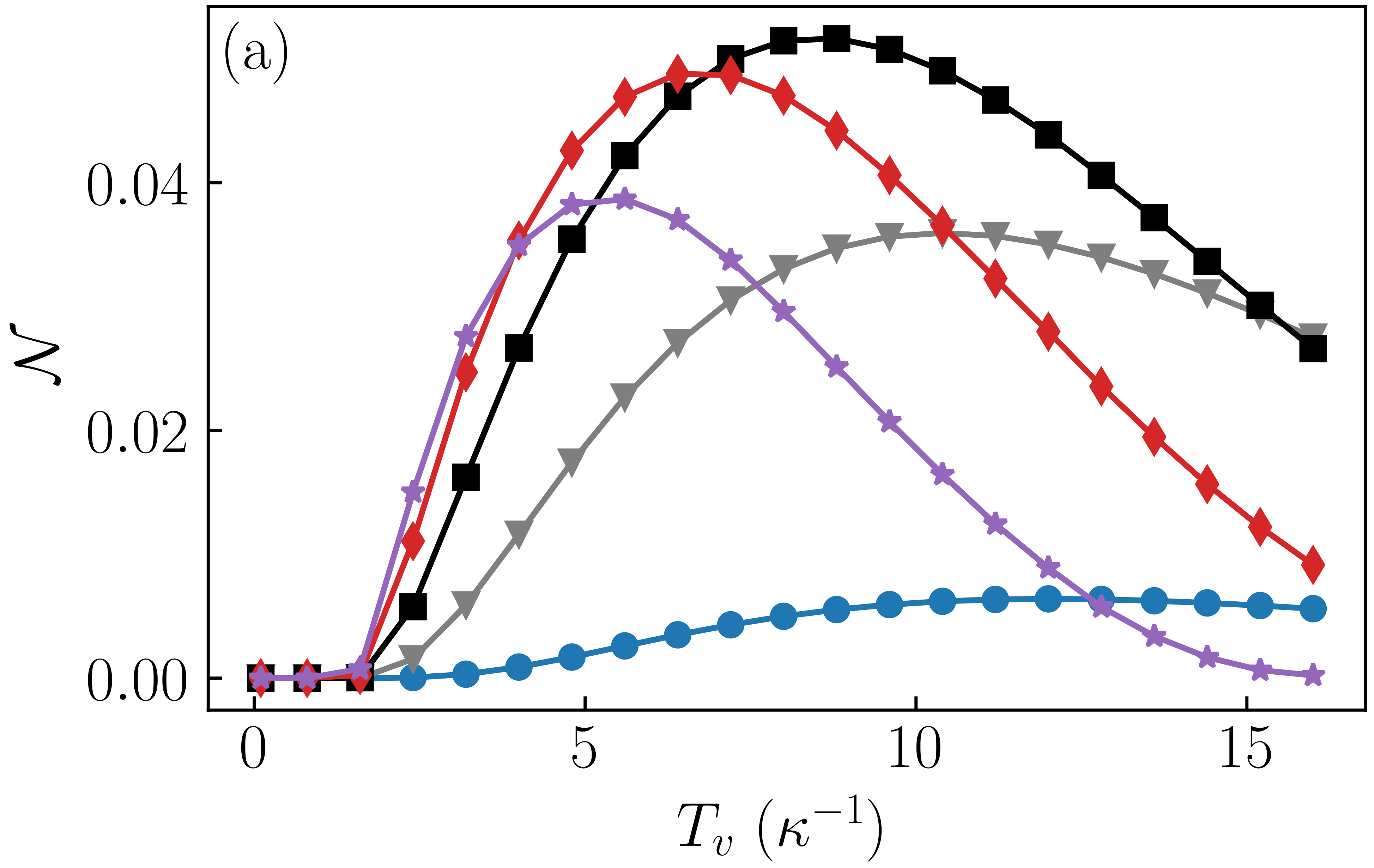} ~ 
  \includegraphics[width=0.23\textwidth]{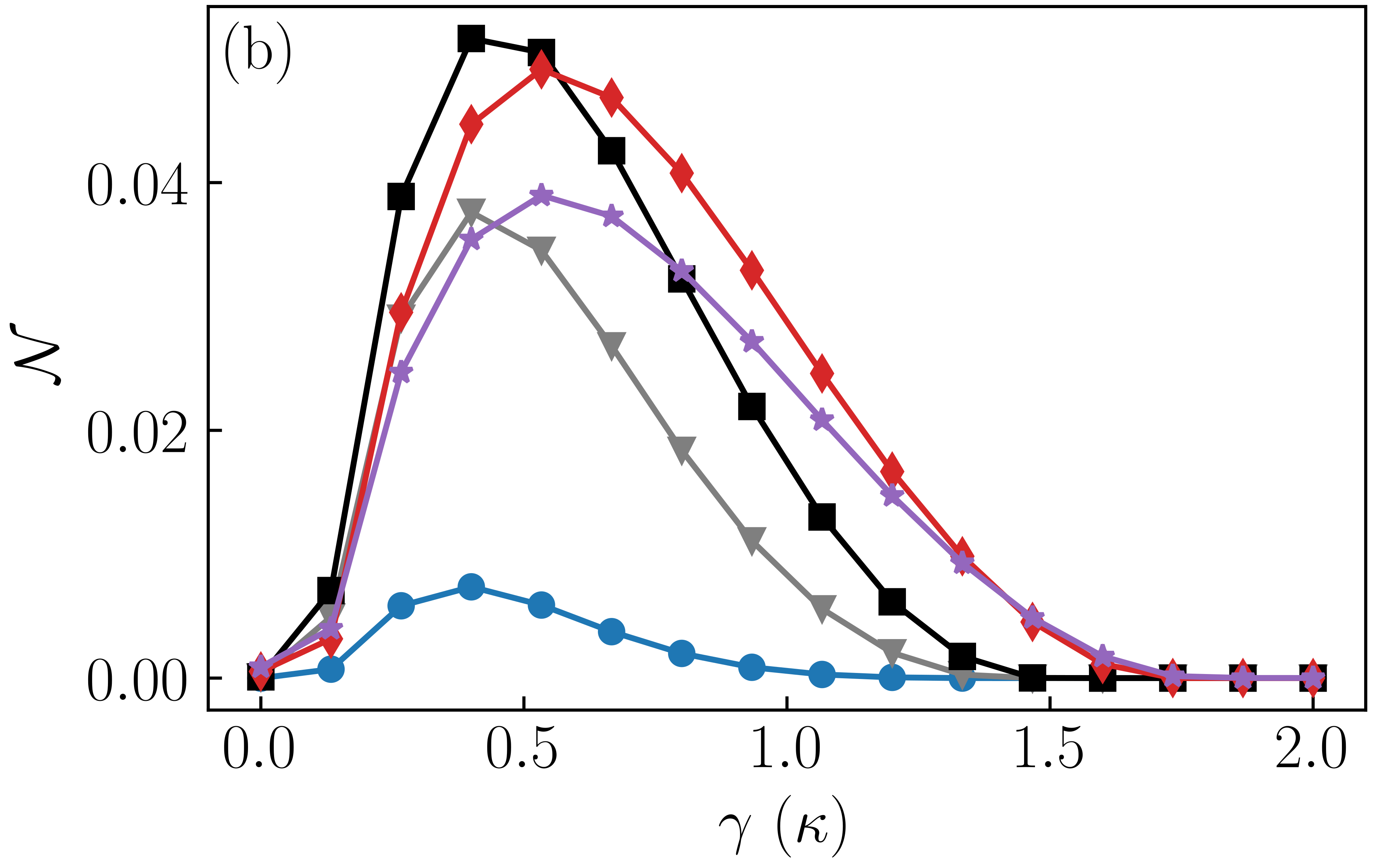}
  \vspace{1mm}
  \centering
  \includegraphics[width=0.23\textwidth]{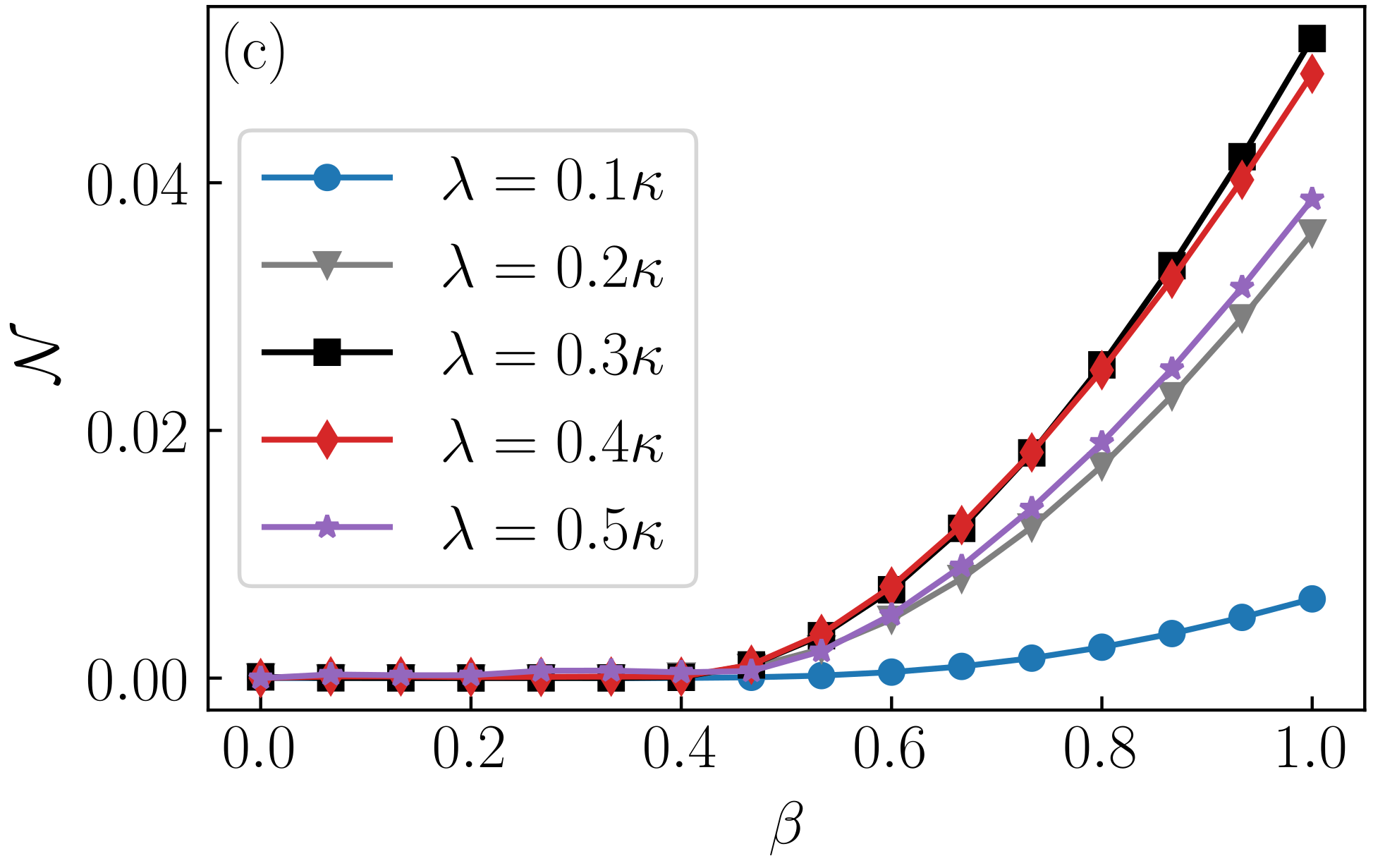}
    \caption{Dependence of the Wigner-negative volume ${\cal N}$ on parameters $T_v$, $\gamma$, and $\beta$ for five different values of the squeezing parameter, $\lambda=\{0.1,0.2,0.3,0.4,0.5\}\kappa$. (a) ${\cal N}$ as a function of the FWHM of the temporal modes, $T_v$. Here, $\gamma=0.5\kappa$ and $\beta =1$. (b) ${\cal N}$ as a function of the decay rate $\gamma$ of the TLS. The FWHM's of the temporal modes are set to the values at which the Wigner-negative volume in (a) is a maximum: $T_v=\{12.0,10.4,8.8,6.4,5.6\}\kappa^{-1}$, respectively, and $\beta=1$. (c) ${\cal N}$ as a function of the coupling efficiency $\beta$ of the TLS to the incident squeezed light, with $T_v$ fixed as in (b), and $\gamma=0.5\kappa$.}
    \label{fig:param_sweeps}
\end{figure}

To quantify the overall extent of negativity in the Wigner distributions we use the Wigner-negative volume \cite{WN_volume}, 
\begin{equation}\label{WN_volume}
    \mathcal{N}=\frac{1}{2}\int \left[ |W(x,y)|-W(x,y)\right] dxdy.
\end{equation}
In Fig.~\ref{fig:param_sweeps} we investigate the dependence of $\mathcal{N}$ on the parameters $T_v$, $\gamma$, and $\beta$. When the FWHM of the temporal modes, $T_v$, is varied (Fig.~\ref{fig:param_sweeps}(a)), we see similar behaviour to the case when the TLS is driven by coherent light \cite{1_atom_2018}. The Wigner-negative volume peaks at a finite value of $T_v$, and this value decreases for stronger squeezing. As $T_v$ increases, more photons are captured in the temporal mode; initially this leads to an increase in Wigner-negativity, as the single- and two-photon contributions become significant. Later, the higher photon number contributions tend to wash out the Wigner-negativity. As the FWHM of the temporal mode is changed, the shape of the Wigner distribution changes as well. Initially in a vacuum state for a zero-width temporal mode function, the central peak of the Wigner distribution gets stretched vertically as $T_v$ increases, as two positive, horizontally stretched, lobes start to form. These lobes then separate vertically and the two negative lobes on either side of the remaining central peak begin to appear and increase in intensity. Eventually, as the two positive lobes 
move further apart vertically, the negative lobes become weaker, and a second set of Wigner-negative lobes appears, as seen in Fig. \ref{fig:4_lobes}. 

In Fig.~\ref{fig:param_sweeps}(b) we vary $\gamma/\kappa$, which amounts to varying the bandwidth of the DPA. For a given $\lambda$ and $T_v$ the maximum negative volume is achieved for a finite value of $\kappa$ on the order of $\gamma$. In the broadband limit, $\kappa\gg\gamma$, a variety of interesting phenomena have been predicted and observed in the spectrum of light emitted by a TLS driven with squeezed light \cite{Sq2A_gardiner,Carmichael1987,Siddiqi2013,Siddiqi2016}, but $\kappa\sim\gamma$ places our system firmly in the regime of non-Markovian dynamics of the TLS, where the characteristic timescales associated with both the DPA and the TLS are comparable \cite{Parkins1988,Ritsch1988,Sq2A_finite}. In this regime, the DPA dynamics cannot be eliminated adiabatically from the model and a simple, atom-only model of the system is not possible. As $\gamma$ is increased, the shape of the Wigner distribution initially follows a similar trend as when $T_v$ is increased, with the central peak separating vertically into two squeezed peaks. However, as the peaks spread apart further vertically, their horizontal width decreases and the negative lobes that had appeared disappear. For very large values of $\gamma$, the Wigner distribution resembles a squeezed state with squeezing along the horizontal quadrature.

Fig.~\ref{fig:param_sweeps}(c) shows that Wigner-negativity exists in temporal modes of the steady-state output field $\hat a_{out}(t)$ even when the coupling of the TLS to the incident squeezed light is not perfect and some of its decay is lost to the greater environment ($\beta <1$). For example, with $\beta\simeq 0.8$ the Wigner-negative volume is still approximately half of that for perfect coupling. When increasing $\beta$ from zero, the squeezing that is initially present in the $y$-quadrature of the output field decreases, as the central positive peak stretches vertically and creates two peaks that begin to separate vertically, allowing for the negative lobes to form on either side of the remaining central peak. These negative lobes become more intense as $\beta$ gets closer to one.
Note that increasing the detuning $\Delta_A$ from zero has a similar effect on ${\cal N}$ as decreasing the coupling efficiency. The shape of the Wigner distribution behaves differently though, as the vertically stretched central peak present for zero detuning rotates to become the horizontally stretched peak, associated with squeezing along the $y$-quadrature, which is present for an infinitely large detuning, where the atom no longer interacts with the radiation emitted by the DPA.

\subsection{Purity and Atomic Excitation}
Temporal modes extracted directly from the output field of a DPA do not show any Wigner-negativity \cite{gz2, Wig_Neg_KPO}; the nonlinearity provided by the TLS is obviously essential to obtaining the negativity seen in the results above. 
It is therefore also interesting to consider the degree of excitation of the TLS, i.e., the population of the excited state of the TLS, $\langle\hat{\sigma}_+\hat{\sigma}_-\rangle$, as a function of system parameters, and how this might relate to properties of the state of the temporal mode, such as the purity.

\begin{figure}[t!]
    \centering    \includegraphics[width=0.46\textwidth]{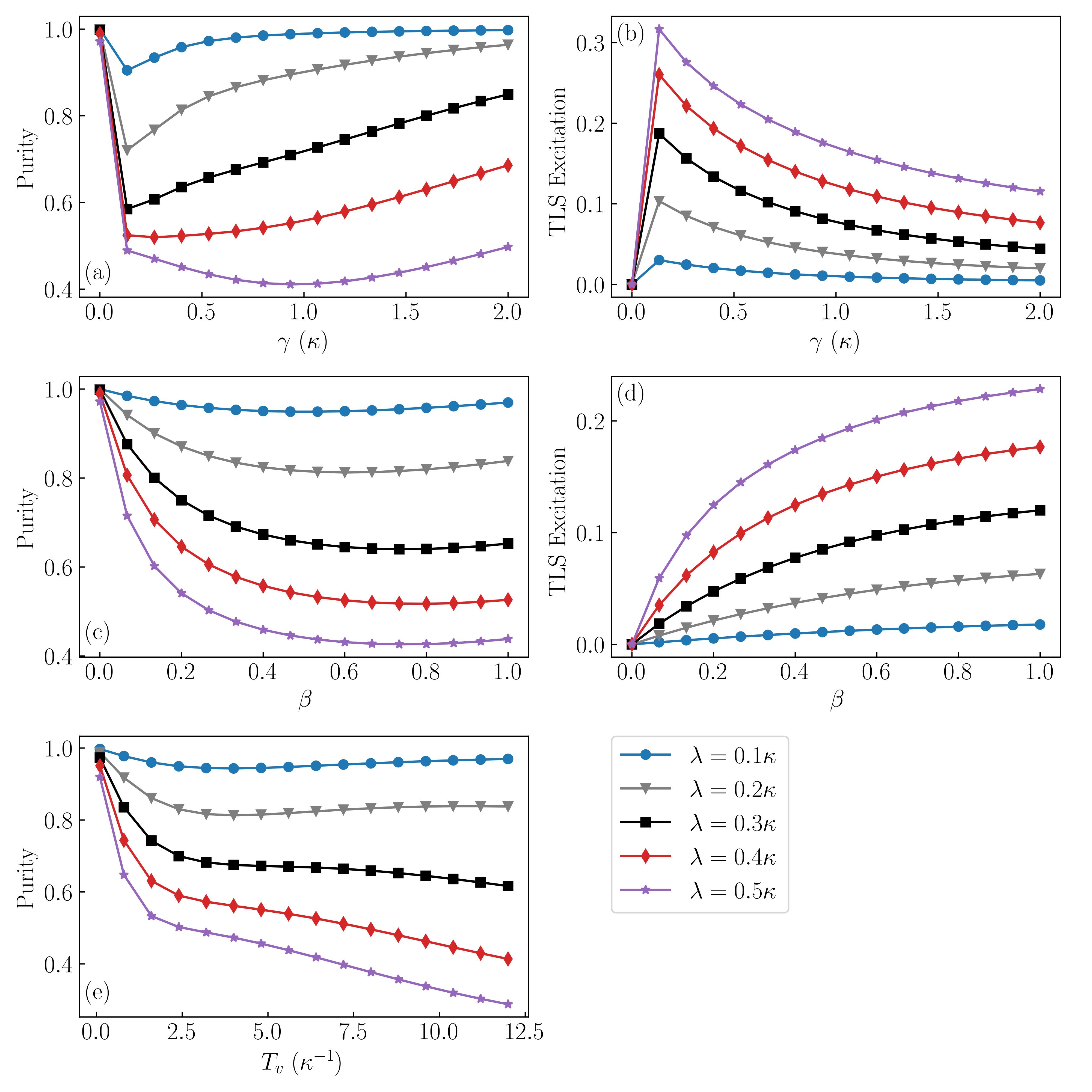}
    \caption{Dependence of the purity of the temporal mode state (left column), and the steady-state excitation of the TLS (right column), on various system parameters. In the first row, (a) and (b) show the dependence of the purity and the TLS excitation, respectively, on the atomic decay rate, $\gamma$, for $\beta=1$. In the second row, (c) and (d) show the dependence on $\beta$, for $\gamma=0.5\kappa$. In the third row, (e) shows the dependence of the purity of the temporal mode state on the FWHM of the temporal mode, $T_v$. The steady-state TLS excitation does not depend on the temporal mode chosen. Each plot shows the behavior for five different values of the squeezing parameter: $\lambda=\{0.1,0.2,0.3,0.4,0.5\}\kappa$. In (a) and (c), $T_v=\{12.0,10.4,8.8,6.4,5.6\}\kappa^{-1}$ for the respective squeezing parameters.}
    \label{fig:purities_atomic_ex}
\end{figure}

In Fig.~\ref{fig:purities_atomic_ex}, we show the dependence of the TLS excitation on various system parameters. Figs.~\ref{fig:purities_atomic_ex}(b) and (d) demonstrate, as expected, that, for a given $\gamma$ and $\beta$, the TLS excitation increases with the degree of squeezing of the driving field (since the intensity of the DPA output field increases with the degree of squeezing, i.e., with $\lambda$). 
Conversely, for a given degree of squeezing, the TLS excitation decreases as $\gamma\, (>0)$ increases (Fig.~\ref{fig:purities_atomic_ex}(b)). Increasing $\gamma$ corresponds to increasing the decay rate of the TLS excitation, so for the same driving strength, a lower steady-state excitation is expected. Note, of course, that for $\gamma=0$ the TLS does not interact with the radiation from the DPA. This is also the case for $\beta=0$, and the TLS excitation increases monotonically towards a maximum as $\beta\rightarrow 1$, where the coupling of the TLS to the DPA output field is maximized (Fig.~\ref{fig:purities_atomic_ex}(d)). 

Figs.~\ref{fig:purities_atomic_ex}(a) and (c) also show the purity of the state of Gaussian temporal mode of the total output field, as given by $\textrm{Tr}\{\hat{\rho}^2_v\}$, as a function of $\gamma$ and $\beta$, respectively. While the purity is not necessarily a monotonic function of these parameters, what is apparent from consideration of these plots and those of (b) and (d) is that the purity decreases as the degree of squeezing and subsequent TLS excitation increase. 
This anticorrelation between purity and TLS excitation is perhaps not surprising, as increased excitation of the TLS can be expected to lead to an increase in the relative contribution of incoherent emission to the total output field and to the extracted temporal mode.

Finally, Fig.~\ref{fig:purities_atomic_ex}(e) shows the dependence of the mode purity on the 
FWHM of the temporal mode. For lower levels of squeezing, the purity is relatively stable against increasing FWHM, but it drops away more significantly for stronger squeezing.


\section{Comparison with squeezed Schr\"odinger cat states.}\label{sec:sq_kittens}
In Fig.~\ref{fig:Wigner_functions}(f) the individual peaks of the Wigner distribution that are displaced from the origin show squeezing in the $\hat{Y}$ quadrature, as confirmed by comparison of the $e^{-1/2}$ contour line with that of a coherent state. We are thus led to consider a comparison of the temporal mode states with a {\em squeezed Schr\"odinger cat state}, or, equivalently, a superposition of displaced squeezed states (SDSS),
\begin{equation}\label{eq:SDSS}
    |\psi_e\rangle=N\left(\hat{D}(\alpha)+\hat{D}(-\alpha)\right)\hat{S}(r)|0\rangle,
\end{equation}
where $\hat{D}(\alpha)=\exp\left(\alpha\hat{a}^\dagger-\alpha^*\hat{a}\right)$ and $ \hat{S}(r)=\exp [r(\hat{a}^{\dagger^2}-\hat{a}^{2})/2]$ are the displacement and squeezing operators, respectively, and $N$ is a normalization constant. 
\begin{table}[t!]
\centering
\begin{tabular}{|c||c|c|c|c|c|c|}
\hline
$\lambda\ (\kappa)$ & 0.1     & 0.2     & 0.3     & 0.4     & 0.5     & 0.6     \\ \hline
$T_v$ $(\kappa^{-1})$      & 12.0    & 10.4    & 8.8     & 6.4     & 5.6     & 5.0     \\ \hline\hline
$\alpha$  & $0.58i$ & $0.74i$ & $0.81i$ & $0.80i$ & $0.81i$ & $0.84i$ \\ \hline
$r$       & -0.06   & -0.14   & -0.25   & -0.42   & -0.58   & -0.83   \\ \hline\hline
$F$  & 0.992   & 0.955   & 0.890   & 0.825   & 0.763   & 0.708   \\ \hline
\end{tabular}
\caption{The values for $\alpha$ and $r$ in Eq.~(\ref{eq:SDSS}) that maximise the fidelity with the temporal mode states captured in the output field $\hat a_{out}(t)$ for given values of $\lambda$ and $T_v$, where $\gamma=0.5\kappa$ and $\beta=1$. The Wigner distributions for the states corresponding to $\lambda=0.1,\ 0.2$ and 0.3 are shown in Fig.~\ref{fig:SDSS}(a), (b) and (c), respectively.}
\label{Tab:kittens}
\end{table}
\begin{figure}[b!]
    \centering
    \includegraphics[width=0.45\textwidth]{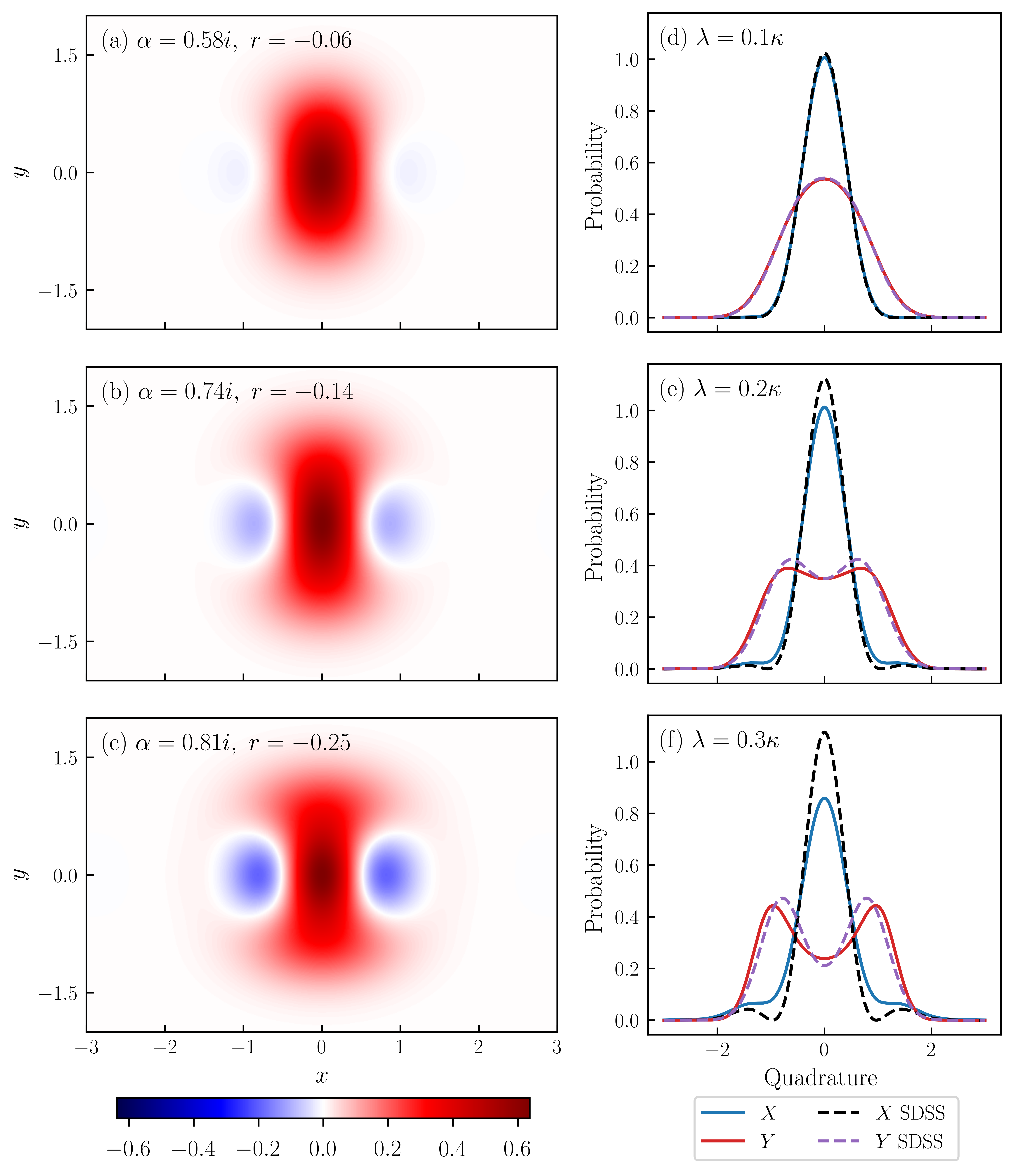}
    \caption{(a-c) Wigner distributions of superpositions of displaced squeezed states (SDSS) which have the highest fidelities with the temporal mode states for which the Wigner distributions are displayed in Figs.~\ref{fig:Wigner_functions}(a), (b), and (c), respectively. Note the different scales of the colour bars in this figure compared to Fig.~\ref{fig:Wigner_functions}. (d-f) Comparison of quadrature probability distributions $\langle x|\hat{\rho}|x\rangle$ and $\langle y|\hat{\rho}|y\rangle$ for the temporal mode states in Figs.~\ref{fig:Wigner_functions}(a-c) and the SDSS states in (a-c) of this figure.}
        \label{fig:SDSS}
\end{figure}
To quantify this comparison we use the fidelity \cite{Fidelity},
\begin{equation}\label{eq:fidelity}  F(\hat{\rho}_1,\hat{\rho}_2)=\left(Tr\left\{\sqrt{\sqrt{\hat{\rho}}_1\hat{\rho}_2\sqrt{\hat{\rho}_1}}\right\}\right)^2.
\end{equation}
For each of the temporal mode states depicted in Fig.~\ref{fig:Wigner_functions} the values of $\alpha$ and $r$ that optimize the fidelity with a SDSS are shown in Table~\ref{Tab:kittens}. For smaller squeezing strengths there is very good agreement. This can also be seen by comparing the SDSS Wigner distributions in Figs.~\ref{fig:SDSS}(a-c) with those in Figs.~\ref{fig:Wigner_functions}(a-c), and from the comparison of the quadrature probability distributions shown in Figs.~\ref{fig:SDSS}(d-f). 
As the squeezing strength is increased the optimal fidelity decreases, but this is not surprising as for stronger squeezing the purity of the temporal mode states decreases (see Fig. \ref{fig:purities_atomic_ex}), plus there is a more significant single photon population, which is completely absent from the idealized SDSS, which only contains even Fock states.

\begin{figure}[b!]
    \centering
    \includegraphics[width=0.45\textwidth]{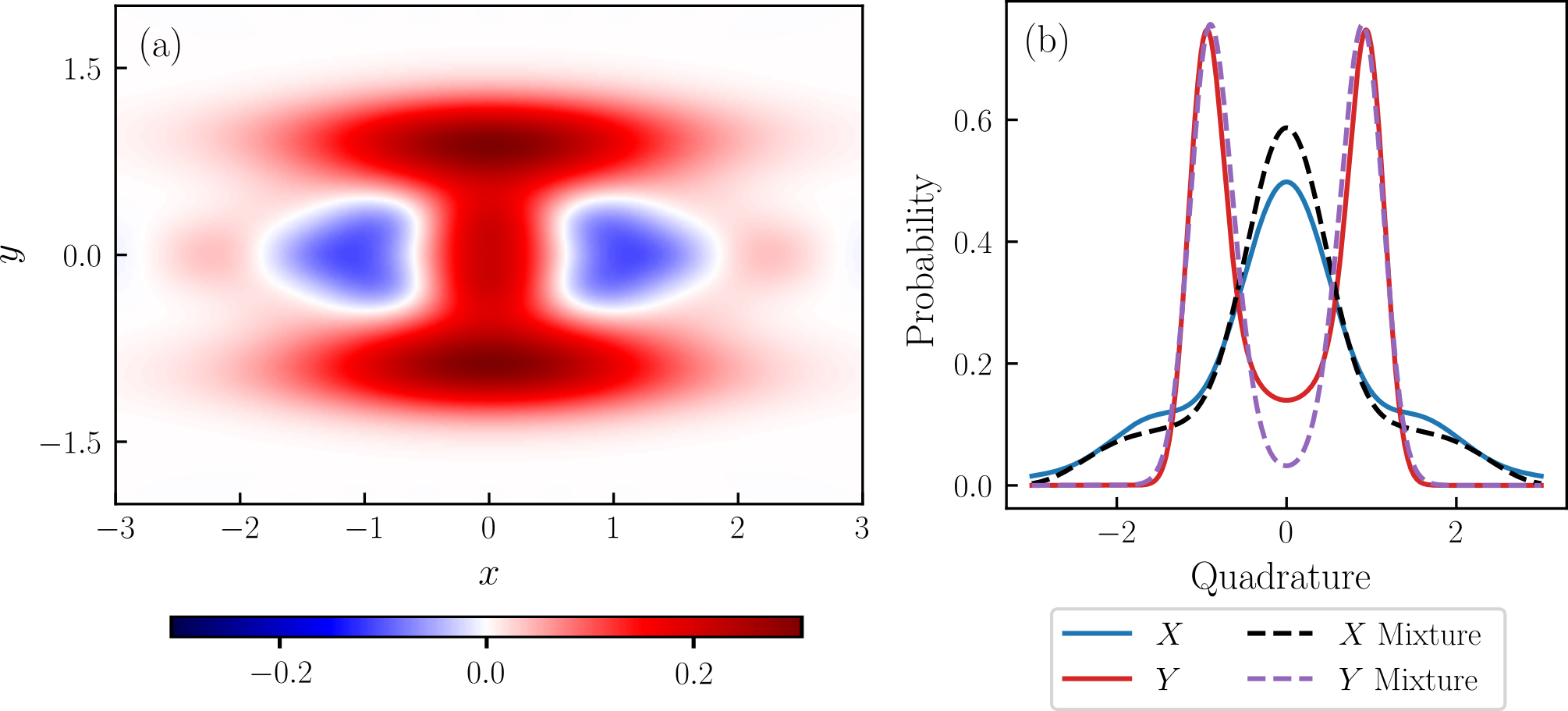}
    \caption{(a) Wigner distribution for the mixed state $\hat{\rho}_M$, with $\alpha=0.81i$, $r=-0.58$, $\alpha_o=0.96i$, $r_o=-0.90$ and $p=0.67$. These parameters were chosen to maximize the fidelity with the temporal mode state depicted in Fig.~\ref{fig:Wigner_functions}(e). (b) shows a direct comparison of the quadrature probability distributions $\langle x|\hat{\rho}|x\rangle$ and $\langle y|\hat{\rho}|y\rangle$ for this temporal mode state and the mixed state depicted in (a). Note that the level of agreement between the quadrature probability distributions is similar to that seen in Fig.~\ref{fig:SDSS}(f) despite this being for a temporal mode state with a significantly larger squeezing strength.}
        \label{fig:mixture}
\end{figure}
We can also consider a comparison between the temporal mode states and a mixture of odd and even superpositions of displaced squeezed states, of the form
\begin{equation}\label{eq:mixture}
    \hat{\rho}_M=p|\psi_e\rangle\langle\psi_e|+(1-p)|\psi_o\rangle\langle\psi_o|,
\end{equation}
where the even component, $|\psi_e\rangle$, is given by Eq.~(\ref{eq:SDSS}), the odd component, $|\psi_o\rangle$, is similarly given by
\begin{equation}\label{eq:odd}
    |\psi_o\rangle=N\left(\hat{D}(\alpha_o)-\hat{D}(-\alpha_o)\right)\hat{S}(r_o)|0\rangle ,
\end{equation}
and $0\leq p\leq 1$. 
This comparison is inspired by the Hinton plots in Fig.~\ref{fig:hinton}, as with this mixed state we include odd photon states without introducing coherences between Fock states differing by an odd number of photons. As the even and odd components in the temporal mode states are completely independent, we can optimize the fidelity of $|\psi_e\rangle$ and $|\psi_o\rangle$ with the temporal mode state separately, before optimizing the probability $p$ to give a maximum fidelity between the temporal mode state and the mixed state $\hat{\rho}_M$. These mixed states in fact  approximate the temporal mode states better that the SDSS, and we can obtain noticeably improved fidelities, especially for larger squeezing strengths. In Fig.~\ref{fig:mixture}, we show the Wigner distribution for the $\hat{\rho}_M$ that optimizes the fidelity with the temporal mode state shown in Fig.~\ref{fig:Wigner_functions}(e), together with a comparison of the $x$- and $y$-quadrature probability distributions for these two states. 
The fidelity between the states is $F=0.932$, significantly improved from the optimum for the SDSS (i.e., for $|\psi_e\rangle$ alone).

Nevertheless, the simplicity of the pure SDSS states is appealing, and the nature of the Wigner distributions and the fidelities obtained establish a strong link between the temporal mode states produced by our system and squeezed Schr\"odinger cat states, especially for small squeezing strengths. Such states are of considerable current interest in quantum optics for the purposes of quantum error correction \cite{SC_code}, especially as they have been identified as a way of realising approximate  Gottesman-Kitaev-Preskill (GKP) states in propagating wave systems \cite{GKP_states, GKP_states_2,GKP_older,F.Squeezing_subtraction,GKP}. GKP states are a particular class of Wigner-negative states that are a promising candidate for quantum error correction in continuous variable computing \cite{GKP_states,GKP_states_2,GKP_older}.

Generating squeezed Schr\"odinger cat states can in principle be done using conditional or heralded schemes based upon the subtraction of photons from a squeezed field using either a beam splitter \cite{GKP_states,BS_kitten,2photons} or a two-level emitter \cite{lund2024subtraction}. A deterministic method has also been proposed \cite{GKP_states_2}, but the scheme involves a large Fock state as input and multiple photon number measurements.
The passive, steady-state nature of our system makes it very appealing for further detailed investigation in this context; in particular, to establish the precise mechanism behind the dynamics that produce the observed transformation of the incident squeezed light. Related numerical simulations have also demonstrated the same effective transformation, and in fact with noticeably enhanced production of Wigner-negativity, for incident (Gaussian) {\em pulses} of squeezed light \cite{Rory}. We note that, although the amplitudes ($|\alpha|$) and squeezing parameters ($r$) of the SDSS considered in the comparisons with our output states are smaller than those needed for many applications \cite{GKP_states_2, SC_code}, it is possible to increase the amplitude of the cat states through conditional breeding \cite{kitten1}, and to apply additional squeezing to the overall state to increase the squeezing parameter.

\section{Conclusion}\label{sec:conclusion}
In summary, we have numerically calculated the Wigner distributions for temporal modes in the steady-state backwards emission of a two-level system driven by finite-bandwidth squeezed light, showing that pronounced Wigner-negativity exists in these temporal modes for a wide range of system parameters. Furthermore, the states that are produced approximate the very topical squeezed Schr\"odinger cat states. The scheme is straightforward and entirely deterministic, and should be achievable experimentally in circuit QED, where efficient driving of a two-level emitter with squeezed light has already been demonstrated \cite{Siddiqi2013,Siddiqi2016}, as has the reconstruction of the Wigner function of a temporal mode in the steady-state output field of a driven qubit \cite{Wig_Neg_Expt}.

\section{Acknowledgements.}
The authors wish to acknowledge the use of New Zealand eScience Infrastructure (NeSI) high performance computing facilities, consulting support and/or training services as part of this research. New Zealand’s national facilities are provided by NeSI and funded jointly by NeSI’s collaborator institutions and through the Ministry of Business, Innovation \& Employment’s Research Infrastructure programme. URL: \url{https://www.nesi.org.nz}.


\bibliography{refs.bib}\label{sec:refs}

\begin{thebibliography}{46}
\bibitem [{dBf()}]{dBformula}%
  \BibitemOpen
  \href@noop {} {}\bibinfo {note} {The degree of squeezing on resonance in the
  DPA output field is given in dB by $-10\log_{10}[1-4\lambda\kappa
  /(\kappa+\lambda)^2]$ (for $\lambda <\kappa$).}\BibitemShut {Stop}%
\end{thebibliography}

\end{document}